\documentclass{aa}

\usepackage{graphicx}

\usepackage{txfonts}

\usepackage{natbib} 
\bibpunct{(}{)}{;}{a}{}{,}

\usepackage{verbatim} 

\usepackage{hyperref}  

\hypersetup{colorlinks=true,linkcolor=blue,citecolor=blue,filecolor=blue,urlcolor=blue}

\usepackage{tikz}
\usetikzlibrary{shapes, arrows.meta, positioning, calc} 

\newcommand{\dydz}{\Delta Y/\Delta Z}

\begin{document}

\tikzset{
        block/.style={
                draw,
                fill=cyan!30,
                rectangle,
                minimum height=3em,
                minimum width=6em,
                align=center,
                rounded corners
        },
        coordinate node/.style={
                coordinate 
        },
}

   \title{A critical analysis of main-sequence fitting in open clusters to derive the helium-to-metal enrichment ratio $\Delta Y/\Delta Z$
}

   \subtitle{}
  \author{G. Valle \inst{1, 2}, N. Ricci \inst{1},  M. Dell'Omodarme  \inst{1},  P.G. Prada Moroni
        \inst{1,2}, S. Degl'Innocenti \inst{1,2}, S. Cassisi \inst{2,3} 
}
\titlerunning{$\Delta Y /\Delta Z$ from open clusters}
\authorrunning{Valle, G. et al.}

\institute{
        Dipartimento di Fisica "Enrico Fermi'',
        Universit\`a di Pisa, Largo Pontecorvo 3, I-56127, Pisa, Italy
        \and
        INFN,
        Sezione di Pisa, Largo Pontecorvo 3, I-56127, Pisa, Italy
        \and
        INAF-Osservatorio Astronomico d’Abruzzo, Via Mentore Maggini s.n.c., 64100 Teramo, Italy
}

   \offprints{G. Valle, valle@df.unipi.it}

   \date{Received ; accepted }

  \abstract
{}
{
We aim to investigate the feasibility of accurately determining the helium-to-metal enrichment ratio $\dydz$ for open clusters using Gaia DR3 photometry.  }
{
To test the reliability of this calibration, we performed a theoretical investigation using mock open clusters. We generated synthetic photometric data from isochrones calculated by five different stellar evolution codes (FRANEC, PARSEC 1.2s, PARSEC 2.0, BASTI, and MIST), for which the true $\dydz$ is known. We then fitted these mock clusters with two sets of isochrones calculated with the FRANEC code, differing only in the implementation of bolometric corrections (BCs). The analysis focused on the $G$-band absolute magnitude range (4.3 to 6.5 mag) to minimise the impact of poorly constrained physics.
Synthetic clusters were generated at [Fe/H] values from 0.0 to 0.15 dex, for different numbers of populating stars and different levels of photometric uncertainties.
}
{
The Monte Carlo experiments revealed significant and code-dependent biases. Unbiased results were achieved only when the stellar models used for synthetic-cluster generation and fitting were identical. Using identical FRANEC stellar models but different BCs introduced a significant bias of up to 0.6. Furthermore, using different stellar models for synthetic cluster generations resulted in even larger biases: $\dydz$ was underestimated by up to 0.8 for PARSEC target isochrones, while it was overestimated for BASTI and MIST isochrones by up to 0.6 and 1.5, respectively. 
}
{
The magnitude and the inconsistency of these biases strongly suggest that the photometric calibration of $\dydz$ using open clusters is not reliably robust.
} 


   \maketitle

\section{Introduction}\label{sec:intro}

The effort to calibrate stellar models has achieved significant results in improving the accuracy of predictions compared with observations. Nevertheless, several physical phenomena ---such as convective transport, microscopic diffusion, and competing processes--- are still affected by notable uncertainties \citep[see e.g.][]{Viallet2015, Moedas2022}.
Moreover, the assumed chemical composition plays a non-negligible role in the overall uncertainty of stellar model predictions. While absorption lines in stellar spectra allow for a robust determination of the surface metallicity, direct measurements of helium abundance are generally infeasible for stars cooler than approximately 15,000 K due to the lack of observable helium spectral lines. Therefore, stellar models must rely on an assumed initial helium abundance. A common approach is to employ a linear relationship between initial helium abundance, $Y,$ and metallicity, $Z$:
\begin{equation}
        Y = Y_p+\frac{\Delta Y}{\Delta Z} Z, \label{eq:dydz}
\end{equation}  
where $Y_p$ is the primordial helium abundance produced in the Big Bang nucleosynthesis and $\dydz$ is the helium-to-metal enrichment ratio.

The precise value of the helium-to-metal enrichment ratio has been extensively studied. A variety of techniques have been deployed to constrain this parameter, including comparing theoretical isochrones with observational data in the Hertzsprung--Russell (HR) diagram \citep[e.g.][]{pagel98, Casagrande2007, gennaro10, Tognelli2021}; fitting evolutionary tracks to a census of nearby field stars \citep[e.g.][]{jimenez03, Valcarce2013, Ricci2025}; utilising asteroseismic data \citep[e.g.][]{Silva2017, verma2019, nsamba2021}; calibrating from detached, double-lined eclipsing binaries \citep{Ribas2000, Fernandes2012, Valle2024dydz}; developing a standard solar model that accurately replicates the Sun's current luminosity, radius, and surface $Z/X$ ratio \citep[e.g][]{Bahcall1995, Serenelli2010, Valcarce2012, Vinyoles2017, Magg2022, Buldgen2025}; calibrating stellar models against evolved stars, specifically horizontal branch and red giant stars \citep{Renzini1994, Marino2014, Valcarce2016}; and leveraging the properties of galactic and extragalactic H II regions \citep[e.g.][]{Peimbert1974, Pagel1992, Chiappini1994, Peimbert2000, Fukugita2006, Mendez2020, Kurichin2021} or planetary nebulae \citep{Dodorico1976, Chiappini1994, Peimbert1980, Maciel2001}. The results derived from these different methodologies exhibit significant scatter and are susceptible to systematic biases. As an illustration, standard solar models commonly yield $\dydz$ values near to 1, while analyses focusing on evolved stars suggest a range from 2 to 3. Conversely, comparisons involving main-sequence (MS) binary stars and the HR diagram generally point towards values spanning 1 to 3, while fitting MS field stars suggests plausible values from 2 to 5.

We recently investigated the robustness and reliability of calibration using detached eclipsing binaries and MS field stars \citep{Valle2024dydz, Ricci2025}. The results indicated the presence of significant systematic errors that undermine the use of these objects for $\dydz$ calibration.
In this paper, we aim to investigate the reliability of estimates attainable from fitting open-cluster MS using a set of isochrones, specifically adopting Gaia Data Release 3 \citep[DR3;][]{Gaia2021} photometry. We particularly focused on results from young (100–800 Myr) and metal-rich ([Fe/H] > 0) clusters, as this age and metallicity range contain near, well-populated open clusters such as the Pleiades, Hyades, Alpha Persei, and Praesepe \citep[e.g.][]{Dahm2015, Brandt2015, MartinEduardo2018, Gossage2018}.
The major challenge in this calibration is the well-known degeneracy in the photometric space among global metallicity $Z$, initial helium abundance $Y$, and age \citep[see among many][]{Chaboyer1992, pagel98, Castellani1999b, Lebreton2001, Pinsonneault2003, Jimenez2004, An2007, Casagrande2007}. 

This degeneracy makes the effect on the isochrones of a variation of $\dydz$ almost indistinguishable from the one due to the metallicity ([Fe/H]) change, within the limits allowed by observational uncertainty. A primary objective of our investigation is to precisely determine if and how well the exquisite precision of Gaia DR3 photometry and astrometry allows us to lift this degeneracy.

To explore the feasibility of this type of calibration, we chose an approach that offers the best sensitivity. Instead of working directly with observational data, we performed a theoretical investigation of the maximum achievable performance using mock data. This approach offers several advantages, mainly allowing for the firm identification of possible hidden biases and systematic sources of uncertainty.
Unavoidable discrepancies exist between theoretical isochrones and actual cluster data. These arise from two main sources: modelling imperfections and observational contamination.
Modelling imperfections arise from missing or incomplete implementation of physical processes in stellar model computations. A related factor is the non-negligible variability that persists in stellar model computations due to the freedom that stellar modelers have in adopting different input physics within their allowed uncertainties \citep[see e.g.][]{incertezze1, Stancliffe2016}. 
Observational contamination occurs because selecting a sample of MS single stars from observed clusters is a process plagued by the presence of unresolved binaries, peculiar objects, and field stars that blur the cluster MS passing all selection steps.
These problems become more relevant for high-mass MS stars and near the turn-off. In these regions, physical phenomena not yet robustly accounted for in stellar models, such as convective core overshooting or rotation, strongly affect the isochrone morphology. Moreover, these zones are significantly less populated than the low MS. This makes the identification of peculiar objects and unresolved binaries more difficult because the effectiveness of techniques such as iterative sigma clipping ---used to identify bona fide single stars \citep[see e.g.][]{DeMarchi2007, goodness2021, Brandner2023a, Brandner2024}--- strongly depends on having a well-populated colour-magnitude diagram throughout all its parts.

\section{Methods}\label{sec:method}

\subsection{Framework}

\begin{figure}
        \centering      
\begin{tikzpicture}[
        auto,
        node distance=1.1cm, 
        >={Latex}]
        
        \node [block,fill=red!20] (Fit) {\textbf{Fit}\\FRANEC PHOENIX\\FRANEC MARCS};
        
        \node [coordinate node, right=of Fit] (sum) {};
        \node [block, ,fill=green!20, right=of sum] (Input) {\textbf{Input}\\FRANEC PHOENIX\\FRANEC MARCS\\PARSEC 1.2s\\PARSEC 2.0\\BASTI\\MIST};

    \node [coordinate node, below=of sum] (sum2) {};
    
        \node [block, below=of sum2] (MC) {Monte Carlo clusters};
        \node [block, below=of MC, fill=yellow!20] (ML) {Maximum Likelihood\\estimates};
        
        \draw [->] (MC) -- node[name=u] {} (ML);
        
        \draw [->] (Input) |- (MC);
        \draw [->] (Fit) |- (ML);
\end{tikzpicture}
\caption{Block diagram of the adopted framework.  }
\label{fig:framework}
\end{figure}
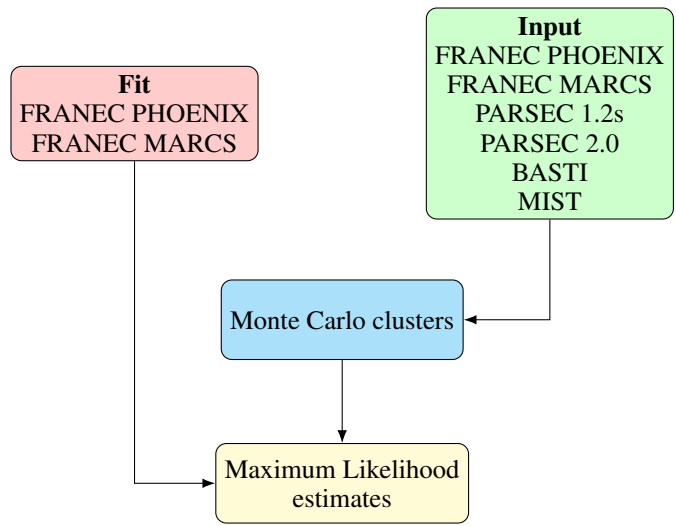

\begin{figure*}
        \centering
        \includegraphics[width=17.0cm,angle=0]{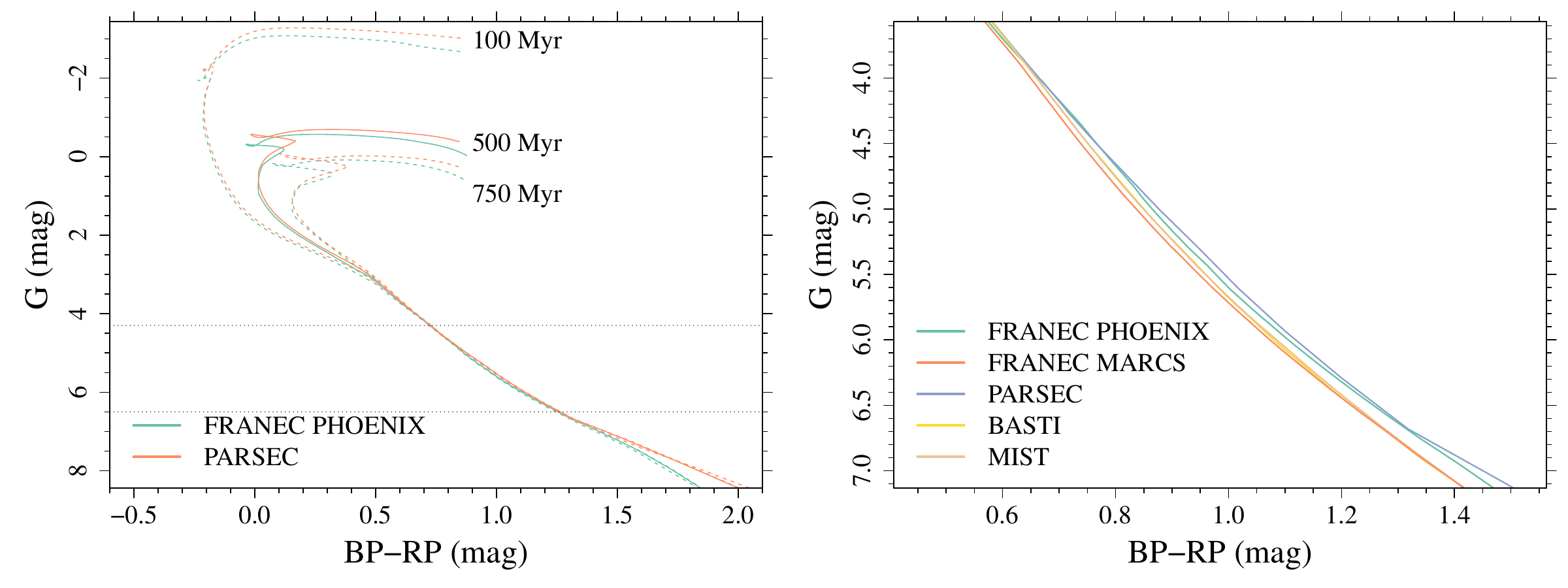}
        \caption{Comparison of isochrones from different stellar evolutionary codes. {\it Left}:
                PARSEC and FRANEC PHOENIX isochrones at [Fe/H] = 0.1 dex and different ages in the Gaia DR3 colour-magnitude diagram. The dotted horizontal lines mark the edges of the selected zone. FRANEC isochrones are computed with $\dydz = 2.0$. {\it Right}: Comparison of isochrones from the FRANEC, PARSEC, BASTI, and MIST codes, at [Fe/H] = 0.1 dex, in the region selected for the analysis.  }
        \label{fig:cmd}
\end{figure*}

To explore the feasibility of reconstructing the $\dydz$ value from mock data, we implemented a structured pipeline. This framework comprises several key components: a grid of fitting models used to estimate the best $\dydz$ value from the synthetic cluster data; a set of isochrones generated from different stellar evolutionary codes; a Monte Carlo procedure designed to generate the synthetic clusters from these isochrones; and, finally, a fitting procedure to obtain the maximum-likelihood estimates of [Fe/H] and $\dydz$. A block diagram illustrating this entire framework is provided in Fig. \ref{fig:framework}. The adoption of target isochrones from various stellar evolutionary codes allowed us to investigate the existence of possible systematic effects in the $\dydz$ estimation within a controlled environment. Discrepancies between stellar models and observations are, in fact, to be expected. Therefore, exploring the impact of these differences ---especially when they originate simply from variations in the morphology of theoretical isochrones--- is of fundamental importance.

Since the various evolutionary sequences along the isochrone are affected by distinct sources of systematic uncertainty, the first step is to identify a portion of the MS suitable for our investigation. We aimed for a region whose morphology is highly consistent among different stellar evolutionary codes and minimally influenced by age changes.
Following \citet{Tognelli2021}, we selected a portion of the isochrones that is most sensitive to changes in $\dydz$. Figure~\ref{fig:cmd} illustrates this by showing a set of isochrones from our fitting grids (FRANEC isochrones) at [Fe/H] = 0.1 dex, $\dydz = 2.0$, and three ages representative of the range considered in this paper. For comparison, the corresponding PARSEC 1.2s isochrones \citep{Bressan2012} are also shown.
The figure demonstrates the impact of age in the high MS region, which begins to manifest itself clearly for $G \lesssim 3.5$ mag. A direct comparison of isochrones at 750 Myr and 100 Myr reveals that, at $G=4.3$ mag, the difference in the $BP-RP$ colour due to age is 0.007 mag for both the FRANEC and PARSEC sets of isochrones. Moving to lower magnitudes causes the impact of age to decrease to a minimum value of $\Delta(BP-RP)=0.004$ mag at $G=5.0$ mag and then increase again, reaching 0.011 mag at $G=6.5$ mag. Similar trends are observed at different metallicities.
Assuming the expected observational uncertainties from Gaia photometry range from 0.005 mag to 0.02 mag (depending on data quality, proximity, and extinction), we safely adopted the range of $G$ = (4.3,6.5) mag for the following analyses.
Enlarging this range is not advisable for theoretical reasons. Specifically, there are well-known discrepancies between isochrones and data at magnitudes below our lower edge \citep{Brandner2023a, Brandner2023b, Wang2025, Ricci2025}. Conversely, stars brighter than the selected upper edge are potentially affected by physical phenomena, such as rotation and convective-core overshooting, whose implementation in stellar evolutionary codes is still an area of active research. Figure~\ref{fig:cmd} shows the differences in the near-turn-off region, where the isochrones from the two stellar evolutionary codes diverge, mainly due to differences in the implementation and efficiency of the convective core overshooting mechanism.

The right panel of Fig.~\ref{fig:cmd} presents a comparison of the corresponding isochrones derived from the different stellar evolutionary codes selected for this analysis; that is, PARSEC 1.2s, PARSEC 2.0, BASTI, and MIST \citep{Bressan2012, Nguyen2025, Hidalgo2018, Choi2016}, as discussed in Sect.~\ref{sec:modelli-target}. The non-negligible spread among the isochrones is immediately apparent, highlighting the relevant impact of adopting different atmosphere models in the computation of the bolometric corrections (BCs). In fact, the difference observed between the two FRANEC isochrones shown here is solely attributable to the choice of alternative BC tables (see Sect.~\ref{sec:modelli}).

\subsection{Stellar model grid for fitting}\label{sec:modelli}

The grid of stellar evolutionary models was calculated for the 0.50 to 4.00 $M_{\sun}$ mass range with a step of 0.025 $M_{\sun}$, spanning the evolutionary stages from the pre-MS to the onset of the red giant branch (RGB).
The initial metallicity [Fe/H] ranged from $-0.5$ dex to 0.35 dex with
a step of 0.01 dex. 
We adopted the solar heavy-element mixture by \citet{AGSS09}. 
For each metallicity, we considered a range of initial helium abundances based on the commonly used linear relation in Eq.~(\ref{eq:dydz}),
with the primordial helium abundance,  $Y_p = 0.2471 \pm 0.001$, from \citet{Planck2020}.
The helium-to-metal enrichment ratio, $\Delta Y/\Delta Z$, was varied
from 0.4 to 3.2 with a step of 0.2. 

The models were computed with the FRANEC code in the same
configuration previously adopted to compute the Pisa Stellar
Evolution Data Base\footnote{\url{http://astro.df.unipi.it/stellar-models/}} 
for low-mass stars \citep{database2012}. The outer boundary conditions  were established by the solar semi-empirical $T(\tau)$  of  \citet{Vernazza1981}, which aptly approximates the results obtained using the hydro-calibrated $T(\tau)$  \citep{Salaris2015, Salaris2018}. 
The models were computed
assuming a mixing-length parameter of $\alpha_{\rm ml} = 2.02, $ which was calibrated by computing the solar standard model.
Atomic diffusion was included by adopting the coefficients given by
\citet{thoul94} for gravitational settling and thermal diffusion. 
To prevent extreme variations in surface chemical abundances, 
the diffusion velocities were
multiplied by a suppression parabolic factor that is equivalent to one for 99\% of the mass of the structure and zero at the base of the atmosphere \citep{Chaboyer2001}.

The raw stellar evolutionary tracks were reduced to a set of isochrones spanning an age range from 100 Myr to 750 Myr, which is the estimated age range of nearby open clusters.
BCs used to derive the Gaia magnitudes were obtained from both the PHOENIX2011 grid \citep{Allard2011} and the MARCS grid \citep{Gustafsson08}.

\subsection{Stellar models for the synthetic clusters construction}\label{sec:modelli-target}

\begin{table*}[ht]
        \centering
        \caption{Relevant input parameters for the stellar evolution codes adopted in this analysis.}
        \label{tab:codici}
        \begin{tabular}{lccccc}
                \hline\hline
        Case & Models   & Mixture & $Y_p$ & $\dydz$ & BC\\
        \hline
        A & FRANEC & Photospheric Asplund 2009 & 0.247 & 1.8 & MARCS \tablefootmark{a}\\
        B & FRANEC & Photospheric Asplund 2009 & 0.247 & 1.8 & PHOENIX \tablefootmark{b}\\
        C,D & PARSEC 1.2s& Caffau 2011 & 0.248 & 1.78 & ATLAS9 ODFNEW, PHOENIX BT-Settl\tablefootmark{c}  \\
        E & PARSEC 2.0& Caffau 2011 & 0.248 & 1.78 & ATLAS9 ODFNEW, PHOENIX BT-Settl \tablefootmark{c} \\
        F & BASTI & Caffau 2011 & 0.247 & 1.31 &   ATLAS9 ODFNEW, PHOENIX BT-Settl \tablefootmark{c}\\
        G & MIST 1.2 &  Proto-solar Asplund 2009 & 0.249 & 1.5 &   C3K \tablefootmark{d}\\
        \hline
        \end{tabular}
        \tablebib{
        (a)~\citet{Gustafsson08}; (b)~\citet{Allard2011}; (c)~\citet{Castelli2003, Allard2012};  (d)~\citet{Conroy2018}.
        }
\end{table*}

To explore the accuracy and precision attainable in a controlled framework, we used several models to build synthetic clusters. We selected isochrones at a fixed age of 500 Myr and metallicities of [Fe/H] = 0.00, 0.05, 0.10, and 0.15 from PARSEC 1.2s \citep{Bressan2012}, PARSEC 2.0 \citep{Nguyen2025}, BASTI \citep{Hidalgo2018}, and MIST \citep{Choi2016}. Isochrones at the desired ages and metallicities were obtained using the respective web interpolator tools.
For PARSEC 2.0 and MIST, we selected models without rotation ($\omega/\omega_c = 0$). For BASTI, we used models computed considering convective core overshooting, but without microscopic diffusion. The target pool also includes the FRANEC MARCS and PHOENIX models, which were sampled at $\dydz = 1.8$.

These target isochrones differ in some key aspects relevant to our investigation. First, they adopt different solar reference mixtures: MIST uses the proto-solar mixture from \citet{AGSS09}, while both the PARSEC and BASTI models employ the mixture from \citet{Caffau2011}. Consequently, the assumed $(Z/X)_{\sun}$ value varies across the different sets of isochrones.
Second, the reference values of $\dydz$ used to compute the models vary significantly. While PARSEC 1.2s and 2.0 adopt $\dydz = 1.78$, BASTI uses $\dydz = 1.31$ and MIST uses $\dydz = 1.5$.
The differences in these inputs affect the $Z$-to-[Fe/H] relations. For example, [Fe/H] = 0.0 corresponds to $Z$ values of 0.01471, 0.01492, and 0.01429 for PARSEC, BASTI, and MIST, respectively. For reference, the $Z$ range corresponding to [Fe/H] = 0.0 of the fitting grid spans the range (0.01266, 0.01329), depending on the value of $\dydz$. 
Finally, the primordial helium abundance values assumed by the different set of isochrones are almost identical (see Table~\ref{tab:codici}); this key point is further discussed in Sect.~\ref{sec:fit-method}.

Since the comparison is performed in the Gaia DR3 magnitude space, another relevant difference relates to the BCs adopted to compute the synthetic photometry. This input plays a key role in isochrone fitting, as shown by \citet{goodness2021}, causing major systematic differences among the calculated magnitudes.
The right panel in Fig.~\ref{fig:cmd} clearly shows the relevance of this input, which causes a large shift between identical FRANEC models. Regarding the significance of this parameter, the displacement in $BP-RP$, resulting from the use of PHOENIX BCs rather than MARCS, roughly corresponds to a difference of 0.15 dex in [Fe/H].
Given this critical importance of the BCs, in the following analysis, we preferred not to anchor the fitting in the target isochrone's [Fe/H] value. Instead, we allowed [Fe/H] to vary as a supplementary fitting parameter. Table \ref{tab:codici} summarises the key inputs adopted by the selected stellar models.

\subsection{Monte Carlo synthetic clusters}\label{sec:MC}

For all the target isochrones discussed above, we generated synthetic clusters by means of Monte Carlo simulations. The simulations were performed assuming three different numbers of stars $(N$ = 50, 100, and 150) ---representing the populations expected within the adopted magnitude range of nearby open clusters---
 and three different levels of photometric uncertainty $(\sigma$ = 0.005, 0.01, and 0.02 mag). Each experiment was repeated ten times, and the estimated parameters were obtained by the mean of the resulting samples.
Therefore, the pipeline steps are as follows: (1) selection of a target isochrone at a given [Fe/H]; (2) random generation of $N$ synthetic stars in the chosen range in the Gaia DR3 magnitudes from this isochrone; (3) random perturbation of the synthetic data adopting Gaussian random errors with a standard deviation of $\sigma$. This intrinsically assumes a synthetic cluster where all the stars to be fitted are effectively single stars, thus implying a perfect rejection of unresolved binaries, field stars, and peculiar objects; (4) fitting of the data using the reference set of isochrones; (5) repetition of steps 1–4 ten times to reduce random errors\footnote{We verified that adopting more repetitions did not modify the results.}.
The repeated experiments were summarised by computing the mean values of the fitted $\dydz$ and [Fe/H].

Overall, we considered the the cases listed below.
\begin{itemize}
\item Case A: Both sampling and reconstruction adopt the same FRANEC MARCS grid. This case serves as our reference and provides the maximum achievable performance in the absence of any systematic discrepancies between synthetic data and models.

\item Case B: Sampling from the FRANEC PHOENIX grid and reconstruction over the FRANEC MARCS grid. This scenario tests the importance of a change in BCs only, since the underlying stellar structure computations are identical.

\item Case C: Sampling from PARSEC 1.2s isochrones and reconstruction over the FRANEC MARCS grid.

\item Case D: Sampling from PARSEC 1.2s isochrones and reconstruction over the FRANEC PHOENIX grid. Comparing Cases C and D allowed us to further investigate the importance of BCs when a systematic discrepancy exists between the synthetic data and the fitting isochrones.

\item Case E: Sampling from PARSEC 2.0 isochrones and reconstruction over the FRANEC MARCS grid.

\item Case F: Sampling from BASTI isochrones and reconstruction over the FRANEC MARCS grid.

\item Case G: Sampling from MIST isochrones and reconstruction over the FRANEC MARCS grid.
\end{itemize}

\subsection{Fitting technique}\label{sec:fit-method}

To perform the stellar fit, we utilised a grid of isochrones computed over a grid of given a enrichment ratio, $\dydz,$ and [Fe/H]. The approach is mathematically equivalent to the standard observational approach of independently fitting $Y$ and $Z$. In the classical framework, the enrichment ratio is typically derived a posteriori by performing a linear regression on the best-fit ($Z$, $Y$) pairs obtained for a sample of clusters. By contrast, our method explores the parameter space treating $\dydz$ as a fundamental input of the grid. 
However, the validity of this equivalence strictly depends on the assumption that the primordial helium abundance, $Y_p,$ is consistent across the different stellar evolution libraries. In fact any offset in the adopted $Y_p$ between the synthetic cluster and the reference grid would introduce a systematic bias in the inferred $\dydz$. Given that the variation of $Y_p$ among the adopted isochrone libraries is only 0.002 (Table~\ref{tab:codici}), the hypothesis of constant initial helium abundance is robust for the purposes of this analysis.

We adopted the fitting technique described in \citet{goodness2021}, which relies on the computation of the Mahalanobis distances between observed data and isochrones. Briefly, we consider a data set consisting of $N$ observations and their associated observational uncertainties. For every $j$-th ($j = 1, \ldots,J$) isochrone in the fitting set, the sum $\chi^2_j$ of the squared Mahalanobis distances, $d,$ between data and isochrones is obtained: 
\begin{equation}
        \chi^2_j = \sum_{i=1}^N d_i^2.
\end{equation}
The likelihood of the $j$-th isochrone is then given by
\begin{equation}
        {\cal L}_j = \exp(-\chi^2_j/2).
\end{equation}
Best-fit values of the metallicity [Fe/H] and $\dydz$ are then computed by a weighted average. Let us define $\theta_j = ({\rm age}, {\rm [Fe/H]}, \dydz)$ as the vector of meta-parameters characterising the isochrone. The best-fitting values are then
\begin{equation}
        \tilde\theta = \frac{\sum_{j=1}^J \theta_j {\cal L}_j}{\sum_{j=1}^J  {\cal L}_j}.
\end{equation} 
The goodness of fit was assessed by means of an $\chi^2$ test, as described in \citet{goodness2021}. For each of the seven considered cases, a Bonferroni correction \citep{Abdi2007} was adopted to protect against Type I errors. Globally, 12 out of 2,520 Monte Carlo simulations ended with no isochrones passing the goodness-of-fit test at level $\sigma$ = 0.05; one passed for case D; two passed for cases B, E, F, and G; and three passed for case C. These simulations were removed from the pool when computing the best-fit values.

\section{Estimated $\dydz$}\label{sec:results}

\begin{figure}
        \centering
        \resizebox{\hsize}{!}{\includegraphics{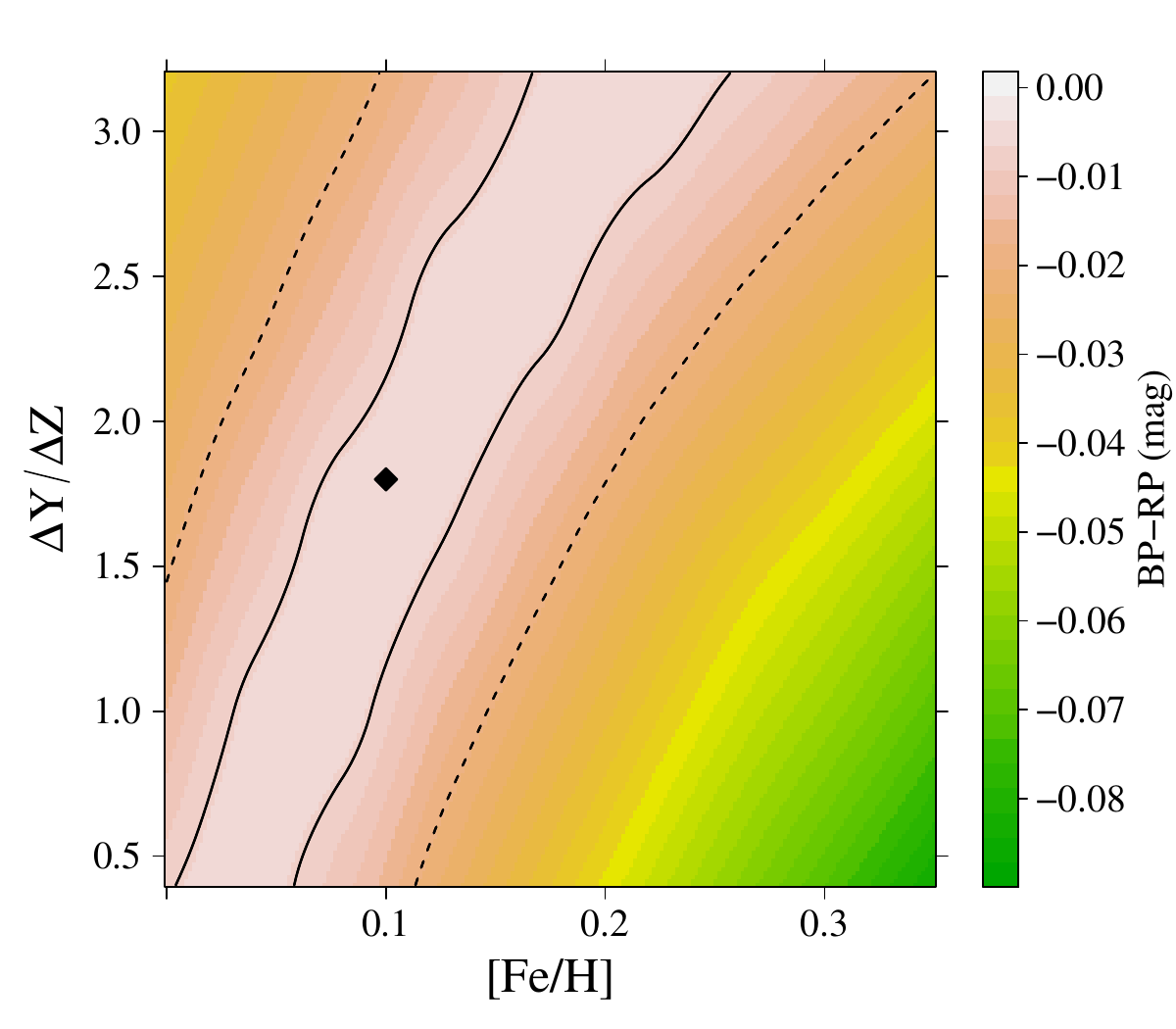}}
        \caption{Filled contour plot of the mean absolute differences in the colour $BP-RP$ between the reference isochrone with [Fe/H] = 0.1 and $\dydz = 1.8,$ and all other isochrones in the grid (for the FRANEC MARCS set). The reference model is identified by the black diamond. Solid and dashed black lines indicate regions where the mean absolute difference is below 0.007 mag and 0.02 mag, respectively.  }
        \label{fig:delta}
\end{figure}

\begin{figure*}
        \centering
        \sidecaption
        \includegraphics[width=12.0cm,angle=0]{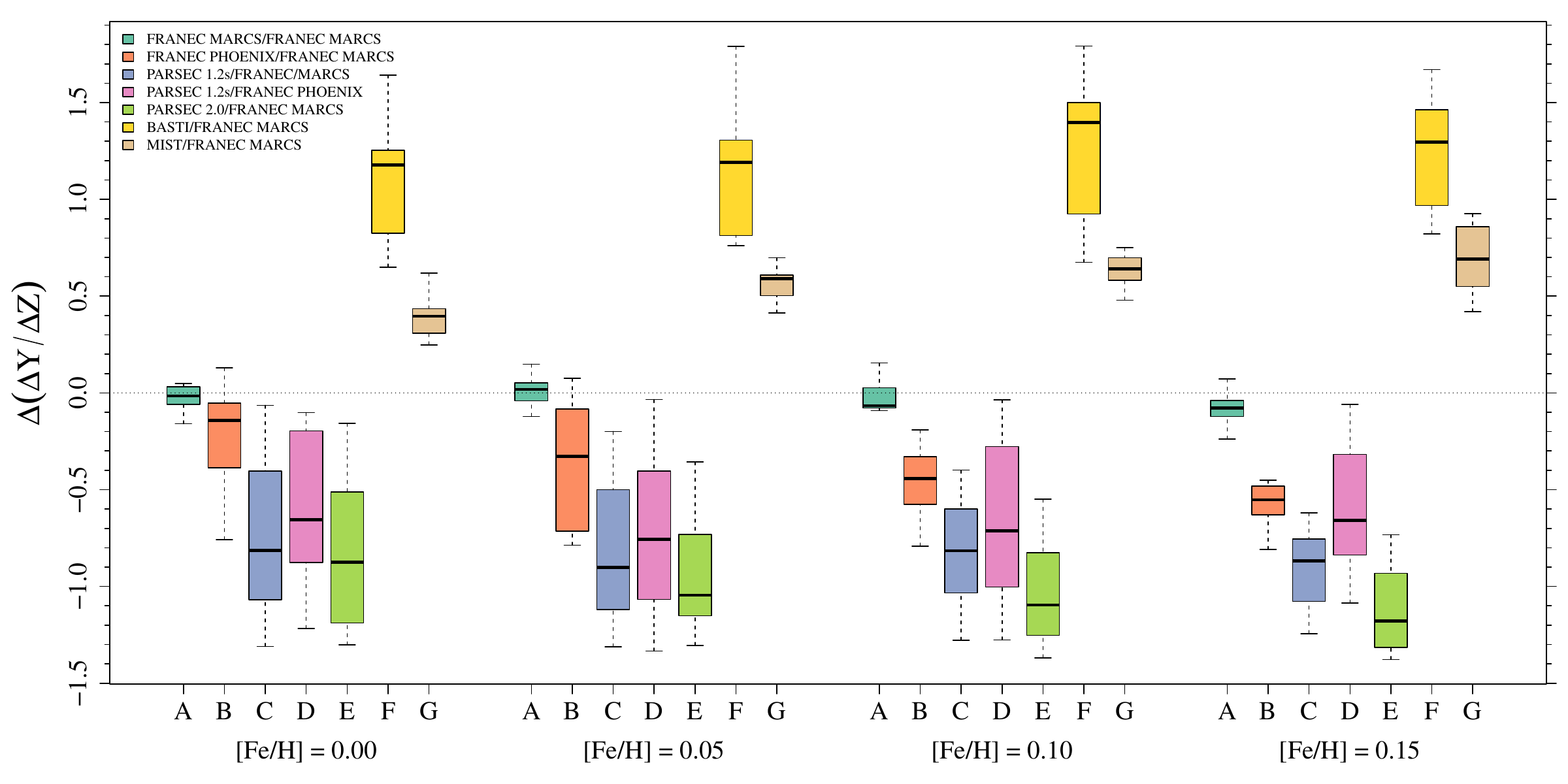}
        \caption{Differences in the estimated values of $\dydz$ for the seven different considered cases with respect to the values adopted by the generating $\dydz$, according to different [Fe/H] values of the target isochrone.   }
        \label{fig:cfr-feh}
\end{figure*}

\begin{figure}
        \centering
                \resizebox{\hsize}{!}{\includegraphics{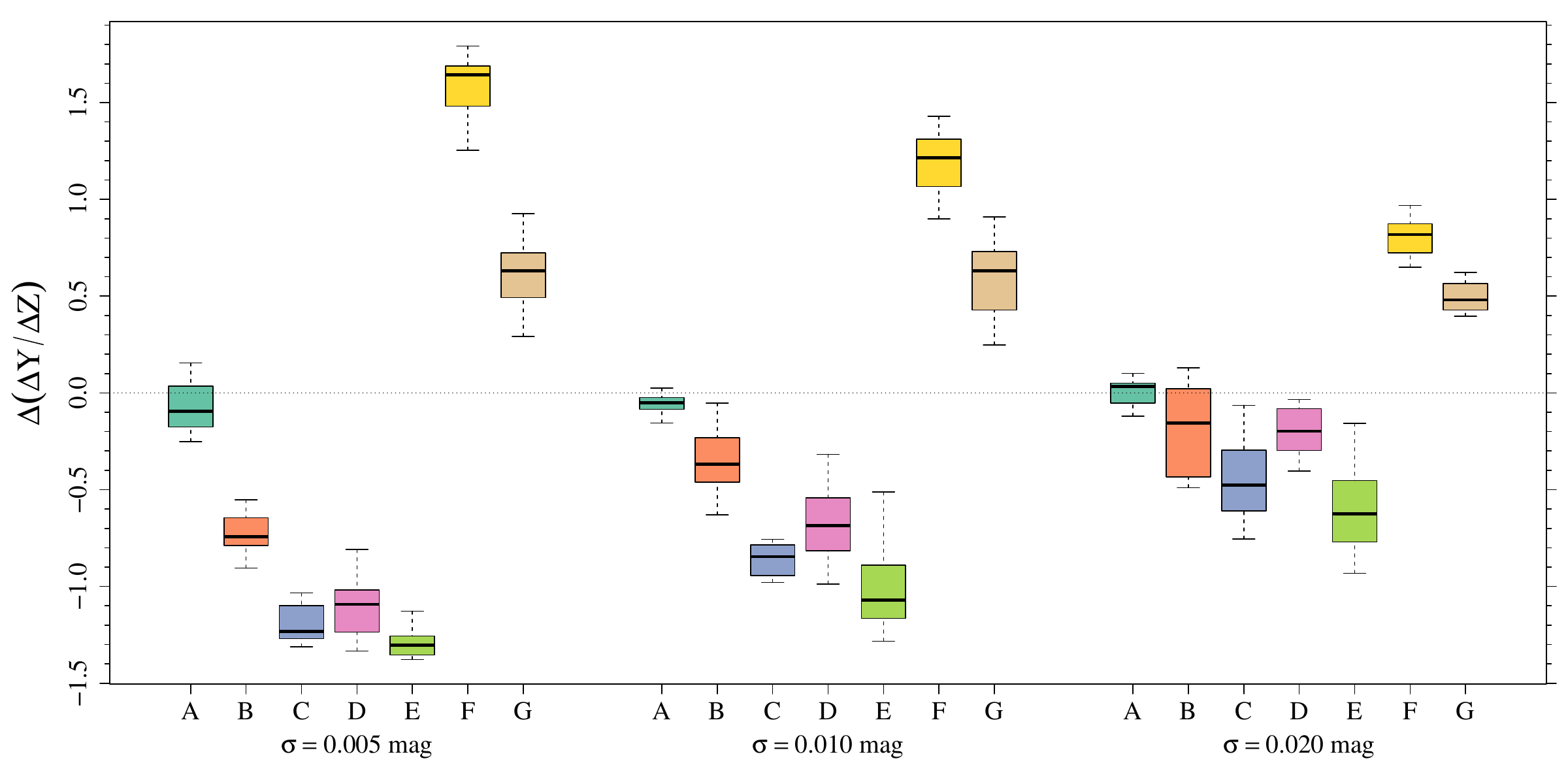}}
        \caption{Same as in Fig.~\ref{fig:cfr-feh}, but according to the different adopted photometric errors.   }
        \label{fig:cfr-err}
\end{figure}

\begin{figure}
        \centering
        \resizebox{\hsize}{!}{\includegraphics{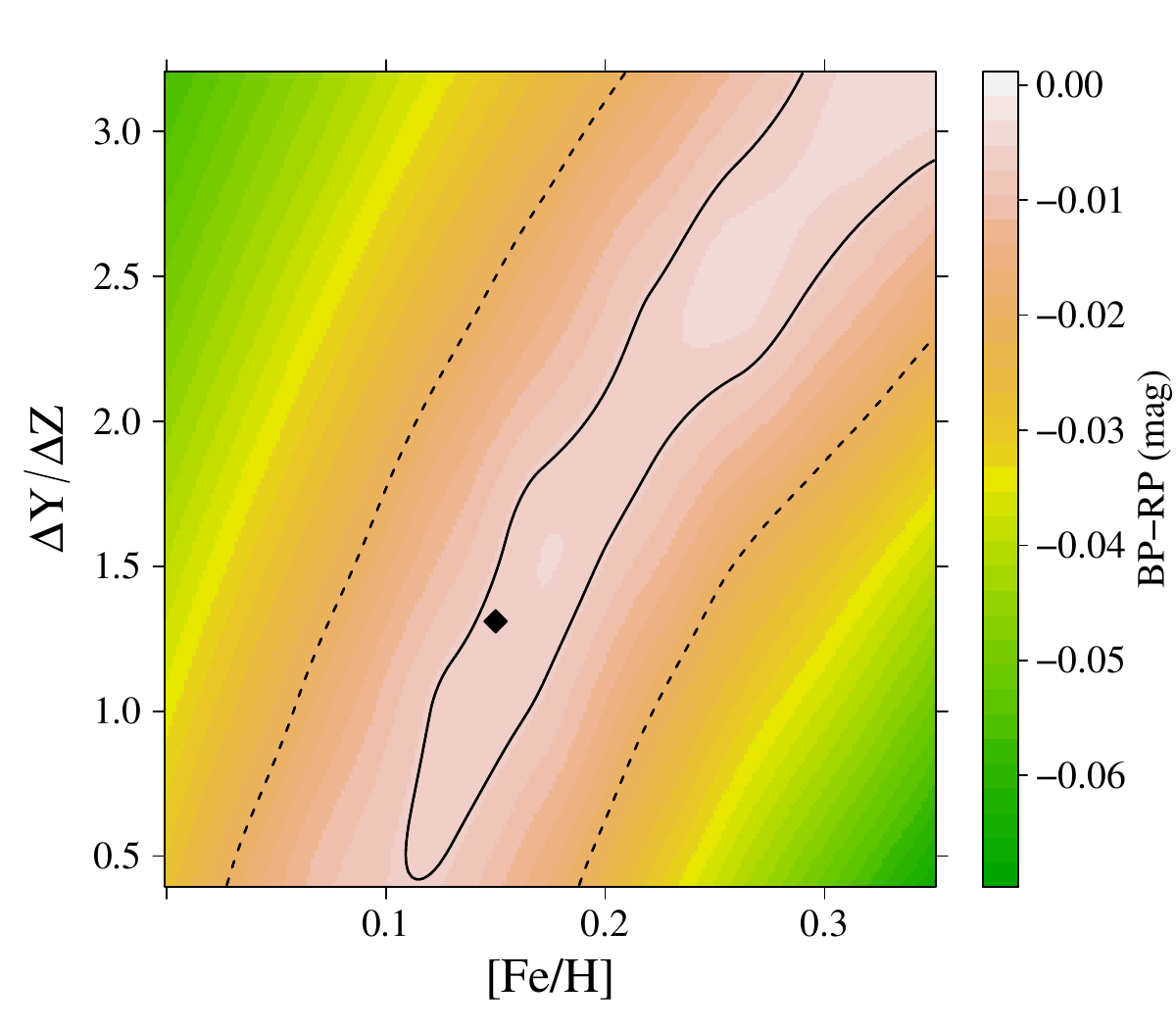}}
        \caption{Same as in Fig.~\ref{fig:delta}, but adopting the BASTI isochrone with [Fe/H] = 0.15 dex as the target.  }
        \label{fig:delta-basti}
\end{figure}

\subsection{Sensitivity of the fitting process}\label{sec:calibr}

As already discussed, the calibration of $\dydz$ is intrinsically a difficult exercise due to the counteracting effect of metallicity and initial helium abundance on the isochrone. This creates a well-known degeneracy. To quantify the relevance of this degeneracy for our purpose, we performed an evaluation over the grid of the differences among the isochrones.
We adopted the FRANEC MARCS isochrone with an age of 500 Myr, [Fe/H] = 0.1, and $\dydz = 1.8$ as our references. We then evaluated the mean absolute difference in the colour $BP-RP$ between this reference and all other isochrones in the grid. This comparison was performed over the magnitude range selected for the analysis, spanning from $G=6.5$ mag to $G=4.3$ mag.
The result of this exercise is shown in Fig.~\ref{fig:delta} as a function of [Fe/H] and $\dydz$. The figure highlights some important points relevant for interpreting the results of this section. First, there is a large region in the metallicity--helium plane, where the difference among isochrones is lower than or comparable to the expected observational uncertainty of 0.007 mag. Second, the boundary values chosen for $\dydz$ in the grid-building process have a fundamental and direct impact on the results. In fact, the figure shows that the zone with low difference extends to both the upper and the lower edges of $\dydz$. Third, even a quite precise determination of [Fe/H], say at 0.05 dex, does not significantly restrict the range of allowed $\dydz$ values.

\subsection{Estimated $\dydz$ and dependence on the target [Fe/H]}\label{sec:feh}

The results of the Monte Carlo simulations confirm that the calibration of $\dydz$ is prone to severe and uncontrolled biases. Figure~\ref{fig:cfr-feh} shows the results of the experiments, grouped according to the metallicity of the isochrone used for the synthetic-cluster generation.
Case A (FRANEC MARCS isochrones for both target and fit) shows that the technique is in principle unbiased in the absence of systematic discrepancies between the data and the models adopted in the fit. The recovered $\dydz$ values are consistent with the value of $\dydz = 1.8$ adopted for the target isochrone, with a remarkable precision of about $\pm 0.2$, independent of the target's metallicity.
For all other cases, the results show a substantial bias. Even in Case B, where the difference between data and models is limited to the BC adopted (PHOENIX for the target and MARCS for the models), the inferred $\dydz$ value is biased towards lower values, and the effect increases with the metallicity. At [Fe/H] = 0.15, the $\dydz$ value is underestimated by as much as 0.6.

The bias is even more severe when the models underlying the synthetic data come from a different stellar evolutionary code. Cases C and D show that the $\dydz$ value estimated when using PARSEC 1.2s isochrones to construct synthetic clusters is largely underestimated when FRANEC models are used for the fit, by about 0.8 for Case C. In agreement with the comparison of Cases A and B, Cases C and D show that the bias is reduced ---by about 0.15--- when FRANEC PHOENIX models are adopted in the fit.
Results from Case E show that the difference between PARSEC 1.2s and 2.0 is quite small up to [Fe/H] = 0.05 dex, but it then substantially increases: at [Fe/H] = 0.15, the $\dydz$ value is underestimated by about 1.2.

The use of BASTI isochrones as targets (Case F) shows an opposite behaviour, as FRANEC MARCS models substantially overestimated the $\dydz$ target value. The overestimation is almost constant, with a positive bias of about 1.3 across the metallicity range and over the target value of 1.3.
Finally, Case G, with synthetic clusters built from MIST isochrones, shows a tendency toward overestimation again, even if it is less severe than Case F. In this case, the estimated value of $\dydz$ increases from 1.9 to 2.2 with increasing metallicity, resulting in a mean overestimation of about 0.6 over the target value of 1.5.

\subsection{Dependence on the other simulation parameters}\label{sec:err}

The dependence on the assumed photometric errors is shown in Fig.~\ref{fig:cfr-err}. It reveals an interesting behaviour that warrants further discussion. As shown in the figure, increasing the photometric errors adopted in both the synthetic cluster generation and the fitting procedure reduces the bias between the target and the estimated $\dydz$. 
However, this reduction occurs by chance and is simply the effect of a regression towards the mean value available in the grid.  
It is likely that the larger photometric errors typical of past observations acted as a statistical smoothing factor, effectively masking the systematic biases that we identified here. 
Figure~\ref{fig:delta-basti} helps us to understand what is happening. This figure is analogous to Fig.~\ref{fig:delta}, which is discussed in Sect.~\ref{sec:calibr}, but the target isochrone comes from the BASTI data set. The differences were computed with respect to the FRANEC MARCS grid.
In this case, the lowest discrepancy region is in the upper part of the allowed $\dydz$ range; therefore, the tendency of Case F fits, discussed above, to prefer $\dydz$ values larger than 2.0 is understandable. When the photometric errors are at the  0.007 mag level, isochrones in this upper region of the [Fe/H] - $\dydz$ plane would provide an acceptable fit. However, when larger photometric errors are allowed, say 0.02 mag, a greater portion of the parameter space must be considered in the fit. As a result, lower values of $\dydz$ become compatible with the data, and the estimated values regress towards the mean $\dydz$ value represented in the fitting grid, which is $\dydz = 1.8$. This behaviour, with increasing observational uncertainty, is not surprising. A similar regression towards the mean value in the grid was previously reported by \citet{Valle2024dydz} in their investigation into the possibility of constraining the $\dydz$ value from MS binary stars.

Regarding the dependence on the sample size, the results align with expectations. The systematic biases generally increase with sample size, with a magnitude that depends on the specific case considered.
In fact, increasing the sample size of the synthetic clusters from 50 to 150 objects limits the impact of random fluctuations, enhances the possibility of detecting the differences in the morphology of the underlying isochrones, and restricts the pool of good fitting isochrones.

The increase in the bias ranges from 0.25 to 0.40 for PARSEC isochrones (Cases C to E), and it is about 0.35 for BASTI (Case F). The effect is more modest (about 0.1) for the comparison between FRANEC models with different BSs and the comparison with MIST models.

\subsection{Restricting to the target [Fe/H]}               

\begin{figure}
        \centering
        \resizebox{\hsize}{!}{\includegraphics{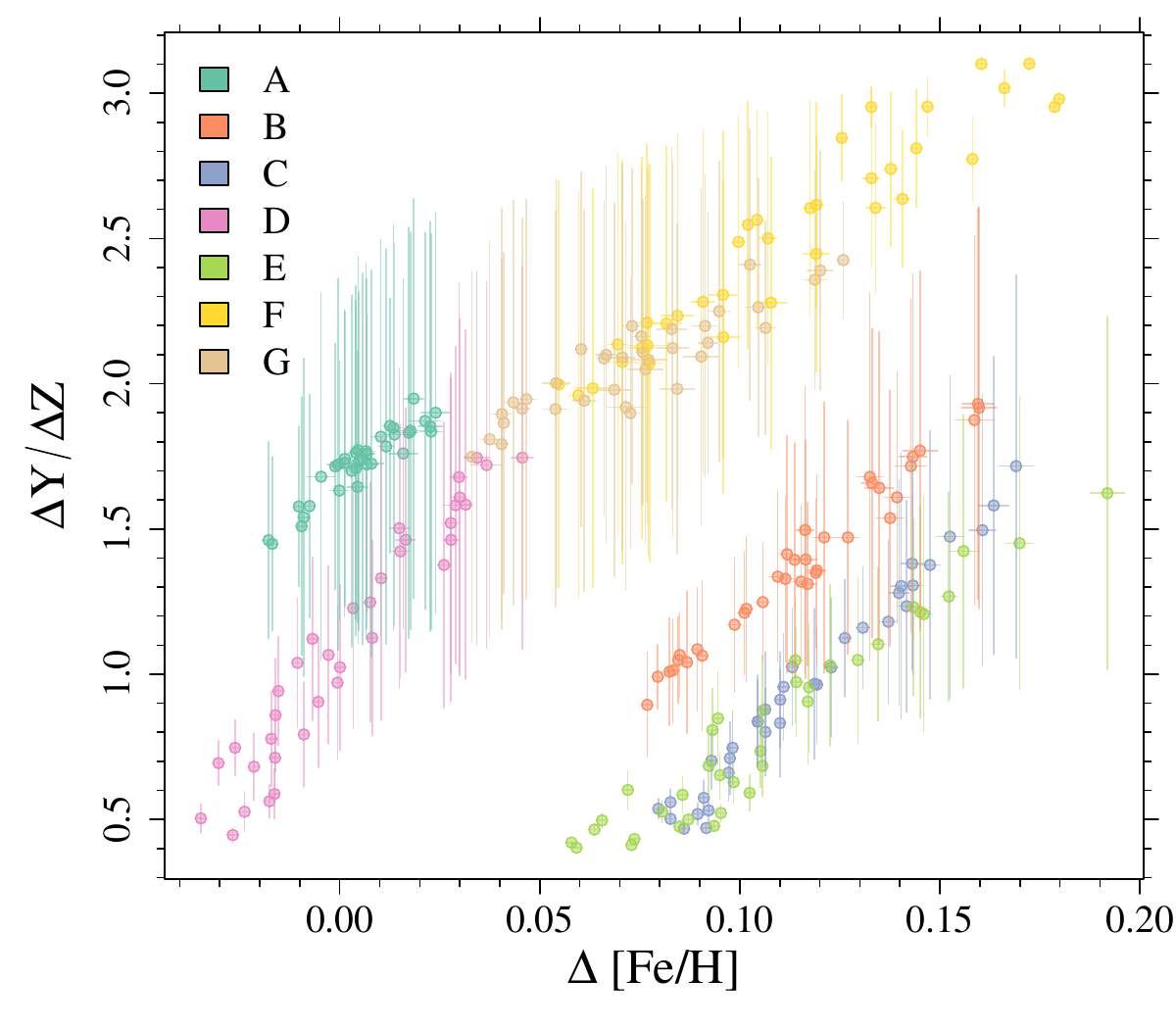}}
        \caption{Scatter plot of the fitted [Fe/H] and $\dydz$ values. The plotted  [Fe/H] are the differences between fitted and target values.   }
        \label{fig:feh-dydz}
\end{figure}

When fitting real stellar clusters, the metallicity [Fe/H] is routinely adopted among the constraints. However, we conducted our simulations lifting this specific constraint, as discussed in Sect.~\ref{sec:modelli-target}. In this section, we explore the impact of imposing a prior in the metallicity.

For this purpose we computed the value $\Delta {\rm[Fe/H]}$ as the difference between the estimated and target metallicities' [Fe/H] for all the discussed simulations. Figure~\ref{fig:feh-dydz} shows the results of our fits in the $\dydz$ versus $\Delta {\rm[Fe/H]}$ plane, grouped by the different explored cases.
It is apparent that only Cases A and D provide a metallicity consistent with the assumed value. The result of Case A is expected because the fitting and target models coincide. In this case, the dispersion of the metallicities is modest, and all the estimated values lie within 0.02 dex of the target. Case D ---fitting target models based on PARSEC 1.2s and reconstruction with FRANEC PHOENIX models--- is also unbiased in [Fe/H], although the dispersion around the true values is substantially larger than in Case A, being about 0.04 dex.

All other cases show a noticeable bias.
Interesting comparisons exist between Cases A and B and between Cases C and D. The former comparison shows a shift of about 0.12 dex in the estimated [Fe/H] for equivalent stellar models that rely upon different BCs. This implies that, as directly verified by means of a dedicated simulation, restricting the fit within 0.05 dex of the value characterising the target isochrones produces no valid output. Increasing the allowed range to 0.10 dex -- which is quite large considering the quoted values for several open clusters -- biases the results toward the low end of the allowed $\dydz$ values. Although the argument is not rigorous, this behaviour can be easily understood by considering the Case B models in Fig.~\ref{fig:feh-dydz}: only a few models are compatible with a shift of 0.10 dex, and these correspond to the low $\dydz$ tail of the distribution. A similar behaviour exists between Cases C and D, which are characterised by a difference in the BCs of the fitting isochrone pool.

\section{Conclusions}\label{sec:conclusions}

We explored the feasibility of constraining the helium-to-metal enrichment ratio, $\dydz,$ using the MS of young open clusters. To this aim, we focused on the Gaia DR3 photometry \citep{Gaia2021}, which provides data of exquisite precision. This method has already been adopted in works relying upon other photometric bands \citep[e.g.][]{pagel98, Casagrande2007, gennaro10, Tognelli2021}, yielding quite different results depending on the models and photometric bands adopted in the calibrations.

To test the reliability of the results that can be obtained with such a calibration, we performed a theoretical investigation. First, following \citet{Tognelli2021}, we identified a region of the cluster MS that is minimally affected by changes in age and is not influenced by still poorly understood input physics, such as stellar rotation or convective core overshooting. This selection restricted the range of absolute $G$ magnitudes to (4.3, 6.5) mag, a range where the differences among various stellar evolutionary codes are lower than in more advanced evolutionary phases.

We investigated the relevance of the morphological differences between data and isochrones on the $\dydz$ calibration by working on mock clusters' data generated from isochrones from a set of different stellar evolutionary codes: PARSEC 1.2s, PARSEC 2.0, BASTI, and MIST \citep{Bressan2012,Nguyen2025, Hidalgo2018, Choi2016}. This setup allowed us to establish the presence of biases in a controlled environment because the target values of $\dydz$ were precisely known. We adopted two different sets of fitting isochrones based on identical stellar models computed with the FRANEC code, but implementing different BCs ---namely PHOENIX and MARCS grids \citep{Allard2011, Gustafsson08}--- to compute synthetic photometry. Synthetic clusters from target isochrones were generated at [Fe/H] values from 0.0 to 0.15 dex for different numbers of populating stars from 50 to 150, and different levels of photometric uncertainties from 0.005 mag to 0.02 mag.

The results of the Monte Carlo experiments evidenced noticeable biases. Only when the synthetic cluster generation and fitting were performed by adopting the same stellar models were the recovered $\dydz$ values unbiased. On the other hand, even adopting underlying identical FRANEC stellar models but different BCs to obtain Gaia magnitudes resulted in biases as high as 0.6 at [Fe/H] = 0.15 with respect to the target value of 1.8.
The biases were even more important when the underlying stellar models were different. While for target PARSEC isochrones the values of $\dydz$ were underestimated by up to 0.8 from the target value of 1.78, opposite behaviour was noted for both BASTI and MIST isochrones. In these cases, we found an overestimation by about 1.3 over the target 1.3 for BASTI and 0.6 over 1.5 for MIST.

An interesting dependence on the magnitude of the assumed photometric error was evidenced. The biases in the $\dydz$ estimation were found to decrease as the photometric error increased. This spurious phenomenon is due to a regression towards the mean value in the fitting grid, which becomes more and more relevant for increasing simulated errors. 
This may explain why, for several decades, it was widely believed that $\dydz$ could be reliably constrained; the lower precision of earlier data sets unintentionally concealed the intrinsic degeneracies of the fits, leading to an overestimation of the method robustness. In the current era of high-precision photometry and dense theoretical grids, the situation has changed. Our results suggest that as observational uncertainties shrink, the underlying systematic biases become the dominant source of error, revealing the fundamental limitations of this estimation technique. The transition from sparse data to the highly accurate observations and massive computational power available today has transformed $\dydz$ from a seemingly accessible parameter into a significant methodological challenge, necessitating a critical re-evaluation of long-standing assumptions in the field.
Actually, when the adopted uncertainty in the simulated data is low, the possibility of discrimination between different isochrone morphology is high.

The occurrence of biases that vary both in direction and magnitude across different tests strongly suggests that the calibration of $\dydz$ using the open cluster MS is fundamentally non-robust. Importantly, this limitation is not a by-product of the specific stellar models adopted here; rather, it reflects an intrinsic sensitivity to the underlying assumptions in the input physics and BCs adopted in the stellar model computations. The fact that these biases scatter significantly when different sets of stellar models are employed ---even when one restricts the comparison to state-of-the art evolutionary codes based on up-to-date and entirely reasonable physical assumptions--- suggests that any alternative analysis conducted with a different code would face the same fundamental limitations.
If a different set of isochrones were used as the fitting pool, the specific values of the biases might shift, but the overall lack of convergence in the results would remain. The inadequacy lies in the method itself: the MS position does not uniquely decouple $\dydz$ from other model uncertainties. Consequently, any calibration of the helium-to-metal enrichment ratio derived from open cluster MS photometry must be viewed with scepticism, as the result is essentially determined by the specific physical input of the chosen model grid and cannot be generalised.

A fundamental limitation identified in our analysis is the significant impact of BC selection on the determination of $\dydz$. Direct evaluation of the mean $BP-RP$ colour difference among the adopted isochrones ---within the $G$ magnitude range selected for this study--- reveals a spread of approximately 0.05 mag. This dispersion significantly exceeds the precision of Gaia photometry for nearby clusters. However, these discrepancies are substantially mitigated in the $G$ versus $T_{\rm eff}$ plane, where the mean spread across the full dataset is reduced to only 20 K. Consequently, performing the $\dydz$ estimation in the $G$ versus $T_{\rm eff}$ plane represents a superior alternative (see e.g. \citealt{Lebreton1999} and \citealt{Casagrande2007} for applications to different photometric systems), provided that high-precision effective temperatures are available.

\begin{acknowledgements}
G.V., P.G.P.M., S.D., and S.C. acknowledge INFN (Iniziativa specifica TAsP) and support from PRIN MIUR2022 Progetto "CHRONOS" (PI: S. Cassisi) finanziato dall'Unione Europea - Next Generation EU. S.C. acknowledges also the support from
INAF - Theory grant "Lasting".
\end{acknowledgements}

\bibliographystyle{aa}
\bibliography{biblio}

\end{document}